\input{aipcheck}

\documentclass[
    ,final            
  ]
  {aipproc}

\layoutstyle{6x9}

\begin{document}

\title{GRB Progenitors and their environment}

\classification{98.70.Rz; 97.60.Bw; 98.38.Hv; 97.10.Me}
\keywords      {gamma-ray bursts; Supernovae; HII regions; 
Mass loss and stellar winds}

\author{Davide Lazzati}{
  address={JILA - University of Colorado, 440 UCB, Boulder, CO
80309-0440, USA} }

\begin{abstract}
The detection of supernova features in the late spectra of several
gamma-ray burst afterglows has shown that at least a fraction of
long-duration gamma-ray bursts are associated to the final
evolutionary stages of massive stars. Such direct observations are
impossible for bursts located at redshift beyond $z\sim0.5$ and
different methods must be used to understand the nature and properties
of their progenitors.  We review the observational evidence for two
particular bursts, for which high quality data are available:
GRB~021004 (at $z=2.323$) and the record redshift GRB~050904 (at
$z=6.29$). We show that both GRBs are likely to be associated to very
massive stars, and that in both cases the progenitor stars were able
to modify their immediate environments with their radiative and
mechanic (wind) luminosity.
\end{abstract}

\maketitle


\section{Introduction}

The quest to understand the progenitor of Gamma-Ray Bursts (GRBs) has
been, and still is, an important branch on the study of these
mysterious astronomical objecs. After many years of struggle and
observations we now know that at least a fraction of the long duration
GRBs are associated to the final evolutionary stages of massive stars
and to type Ic supernova (SN)
explosions\cite{Galama98,Stanek03,Hjorth03,Malesani04,Campana06}.

Even though the association is fairly robust in some cases, all the
strong associations are observed in low redshift GRBs, characterized
most of the times by lower than usual $\gamma$-ray energies. This
could well be a selection effect, since SN features can be observed
only for small-intermediate redshifts, and low redshift events sample
the lower luminosity tail of the luminosity function. Alternative
evidence has come about recently, showing that some long duration
events are not associated to bright
SNe\cite{Dellavalle06,Galyam06,Fynbo06}. For the brightest high
redshift events the issue is therefore open and indirect indicators
must be used to constrain the nature of the progenitors. In addition,
even for the bursts with a robustly associated SN, important questions
like the properties of the progenitor star, the environment, the
metallicity, and the mass loss rate have to be understood.

Traditionally, progenitor indicators are of two classes: the SN hump
and the properties of the immediate GRB environment. At very high
redshift ($z>2$), supernova humps become more and more difficult to
detect and the investigation of the environment is the most promising
method to constrain the progenitor properties. The environment of GRBs
can be studied either with absorption studies or by modelling the
afterglow emission. Afterglow modelling\cite{Panaitescu01} relies on
the spectral and temporal modelling of the afterglow lightcurves to
infer density and structure of the ambient medium at a distance of
$\sim1$~pc from the burster. Even though the technique is in principle
very effective, the diversity of afterglows has, especially after the
advent of {\it Swift}\cite{Obrien06}, made clear that the model is
too simple and the results of the modelling either unreliable of
dependent on the assumptions made.

The study of absorption continua and features in the optical and X-ray
afterglows is potentially a much more powerful probe. Not only it is
sensitive to the structure and density structure of the absorber, but
can also constrain its metallicity, temperature and dust content. The
major problem with absorption measurement is the fact that the
distance between the absorber and the light source is unknown. In the
case of GRBs, the photon flux within several tens of parsecs from the
burster is so large that the status of the absorbing material is
modified in a time-dependent way\cite{Perna98,Lazzati01}. The effect
is generally a reduction of the opacity during the prompt phase and
the early afterglow. The system can be completely solved since the
speed of the absorption decrement depends on the distance from the
burster, while the normalization of the absorption depends on the
total amount of absorber present. In practice, a forward fitting
method is adopted since the modeled quantity is highly convoluted.

\begin{figure}[!t]
\includegraphics[width=0.9\textwidth]{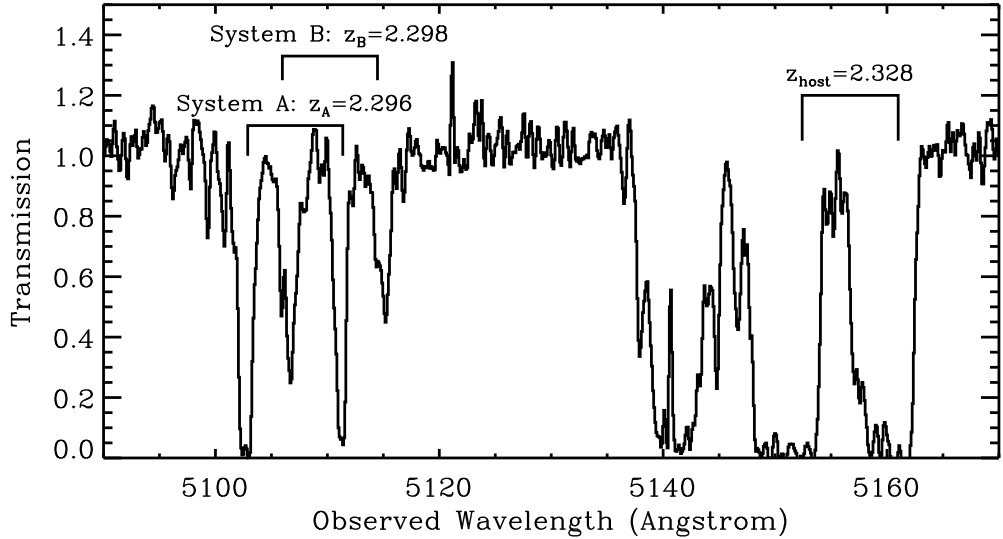}
\caption{{A portion of th spectrum of GRB~021004 from the VLT
data\cite{Fiore05} around the CIV doublet absorption complex.}
\label{fig:spex}}
\end{figure}

In this review, I concentrate mainly of two events, for which high
quality data are available, enabling a thorough modelling of their
environments. The first event is GRB~021004, at redshift
$z=2.323$. The second event is GRB~050904, the highest redshift GRB at
$z=6.29$\cite{Tagliaferri05,Haislip06}.

\section{GRB~021004}

Despite the high redshift, GRB~021004 had a very bright afterglow,
characterized by high variability on top of the smooth power-law
decay\cite{Lazzati02}. It is, to date, the high redshift afterglow
best studied
spectroscopically\cite{Moller02,Mirabal03,Matheson03,Schaefer03,Fiore05,Starling05}.
The observational interest was triggered by the discovery of very high
proper motions in the CIV and SiIV absorption complexes, indicating
that material is outflowing from the GRB site towards the observer at
speed as large as $\sim3000$~km/s\footnote{The possibility of an
intervening system was ruled out by statistical
arguments\cite{Moller02,Schaefer03}. The discovery of H Lyman $\alpha$
and $\beta$ absorption associated to the high speed
absorption\cite{Mirabal03} has however put this conclusion under
debate, at least for one of the absorption systems\cite{Chen06}.}.

Fig.~\ref{fig:spex} shows the CIV absorption around the redshift of
GRB~021004 in the high-resolution VLT data\cite{Fiore05}. Several
absorption systems are visible. We consider here the two systems
labeled $A$ and $B$, since the CIV and SiIV absorptions in the other
systems are heavily saturated and therefore it is not possible to
measure a reliable column density of the absorbing ions. Twelve
measurement of CIV absorption from systems $A$ and $B$ combined are
available, taken at times ranging from $11$~hours to $\sim7$~days
after the event\cite{Lazzati06}. Analogous measurement, albeit at a
lower signal to noise ratio, are available for SiIV. The data show a
moderate evidence (about $2\sigma$) of decrease of the equivalent
width $EW$ in the CIV absorption\cite{Lazzati06}. Thanks to the
availability of high resolution spectroscopy, it is possible to
measure the temperature of the absorbers that turns out to be
approximately of $\sim10^6$~K.

\begin{figure}[!t]
\includegraphics[width=0.7\textwidth]{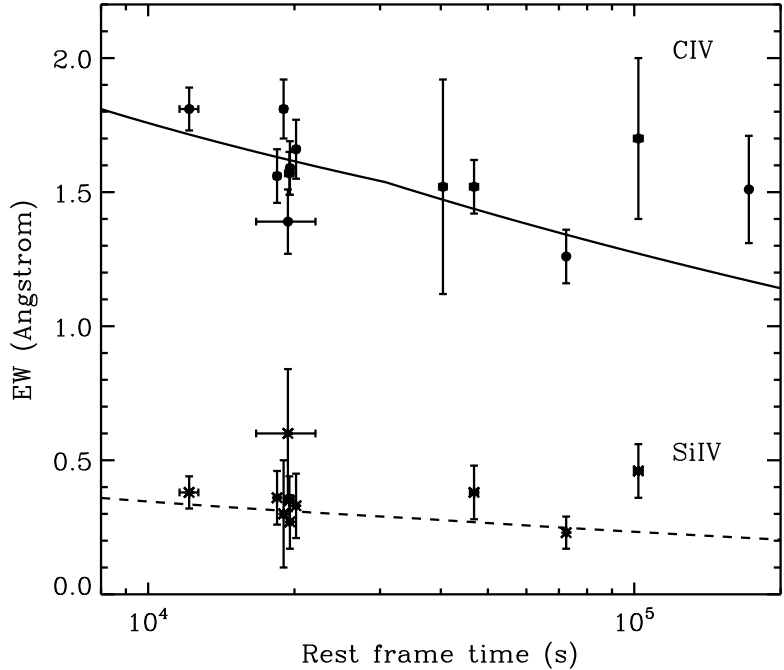}
\caption{{Equivalent width of the CIV and SiIV absorption of systems 
$A$ and $B$ combined. The lines show the best fit evolution of the
lines for absorbing material with a wind distribution out to a radius
of $\sim100$~pc.}
\label{fig:fitc4}}
\end{figure}

The high outflow velocity of the absorber cannot be explained with
motions in the galaxy, since those are expected to be one order of
magnitude smaller. A likely candidate is the high velocity wind
ejected by a Wolf-Rayet star in the WNE or WC phases. Winds are
however usually too cold to display strong high-ionization lines.  The
presence (and possible decay) of CIV absorption therefore allows us to
constrain the distance of the absorber. The carbon (and silica) cannot
be too close to the burst, or it would be completely ionized by the
burst radiation to CV, CVI and CVII. On the other hand, it must be
close enough for the photons to convert most of the carbon into
CIV. Additional constraints come from the EW variability, which is
however marginal for this event.

Figure~\ref{fig:fitc4} shows the result of the fitting of a numerical
photoionization model\cite{Perna02} on the data\cite{Lazzati06}. The
model assumes that the absorbing material is distributed as a wind
with mass loss rate $\dot{M}\sim2\times10^{-4}$~$M_\odot$/y and with a
wind termination radius $R_{\max}>80$~pc. The metallicity assumed
ranges from $100$ to $10000$ times solar. The distance of the wind
termination shock is surprising for at least two reason, but the
interpretation is solid as long as the high velocity features are
considered local to the GRB (see footnote 1).

First, a wind termination shock so distant implies that the external
shock propagated in a stratified medium, while many authors claim the
afterglow properties to be consistent with a uniform ISM. In addition,
should all GRBs be surrounded by analogous winds, all afterglows
should be well described by a wind model, contrary to the results
obtained in the modelling\cite{Panaitescu01}. Second, it is very
unusual that the wind termination shock lies at such a large distance
from the stellar progenitor. This not only requires a very massive
progenitor star, but also a very low density
environment\cite{Dwarkadas05,Eldridge06}.

\subsection{Shielding}

It has been suggested that a clumpy wind with clumps of approximately
$M=10^{-4}M_\odot$ located at $\sim1$~pc from the burster would be
able to shield the CIV from the burst ionizing photons, allowing for a
much smaller termination shock radius\cite{Chen06}. This is not a
viable solution. The time integrated prompt emission and early
afterglow of GRB~021004 had $N_\gamma=4\times10^{60}$ photons. At a
distance of $1$~pc, this corresponds to a flux of $3.3\times10^{22}$
photons per square centimeter. Since the average photoionization
cross-section is approximately $10^{-19}$ to $10^{-18}$~cm$^2$, it can
be easily seen that the clumps will be very quickly ionized (each
electron ``sees'' approximately $3000$ photons capable of ionizing it).

As a consequence, the shield will be completely ionized in a fraction
of a second after the burst onset. By the time the spectroscopic
observations are performed, it will be completely transparent, and the
ionization of carbon will have expanded out to approximately
$100$~pc\cite{Lazzati06}. The proposed shielding effect\cite{Chen06}
is not a solution.

\section{GRB~050904}

GRB~050904 is a test case to show the potential of modelling the
evolution of continuum absorption, in this case in the soft X-rays.
This burst was discovered at a very large redshift $z=6.29$, and was
followed with the {\it Swift} XRT telescope as well as by various
optical telescopes. An optical spectrum was obtained with the Subaru
telescope\cite{Kawai06}. The optical spectrum shows evidence of
absorption from SiII local to the GRB host. From the analysis of
forbidden line ratios, is was derived that the afterglow photons
propagated through a medium with an electron density of several
hundreds\cite{Kawai06}. Such absorber is local to the GRB host galaxy,
but must lie at a distance of at least $\sim10$~pc from the burster in
order for the gas to avoid complete ionization.  An analogously high
density of the close environment of the GRB is instead inferred from
radio observations\cite{Frail06}.

\begin{figure}
\includegraphics[width=0.8\textwidth]{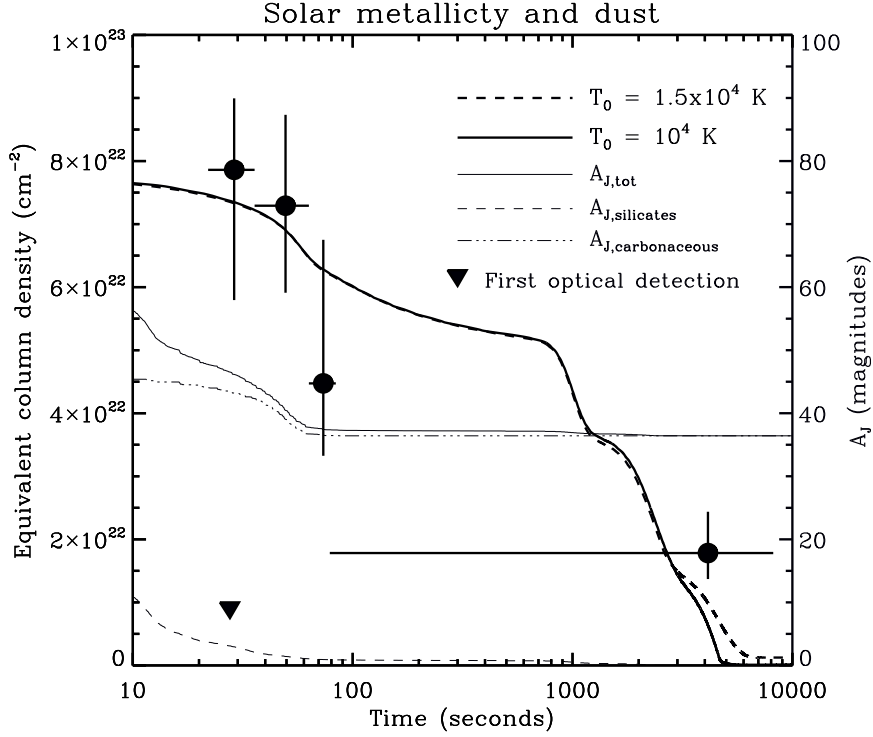}
\caption{{Evolution of the equivalent column density measured in the
X--ray afterglow of GRB~050904 (dots with errors at $1\,\sigma$).
Time is in the rest frame. Solar metallicity with no Fe and Ni has
been assumed. The equivalent column density is defined as the column
density that would produce the same amount of absorption for a cold
non-ionized absorber.  The solid and dashed thick lines show the best
fit models for different initial temperatures. The photoionization
code has in input the observed light curve of GRB~050904. The drop in
absorption at $t\sim1000$ s (in the rest frame) corresponds to the
group of bright X--ray flares.  The thin lines (and right y axis) show
the amount of absorption that would be observed in the J band (rest
frame $\sim 7.2$~eV) if the X-ray absorbing medium would be polluted
with Galactic-like dust. The optical transient was observed at
$t_{\rm{obs}}=200$~s in white light, indicating very little
absorption. Thin dashed and dot-dashed lines show the absorption due
to silicates only and to carbonaceous grains only, respectively. The
little extinction implied by the early optical observation can be
explained by a dust component rich in silicates and depleted in
carbonaceous grains. This could be the results of an ISM enriched by
pair instability SNe.}
\label{fig:nh}}
\end{figure}

Time resolved spectroscopy of the early X-ray afterglow revealed the
presence of soft X-ray absorption in excess of the Galactic
value. Such excess absorption is not constant but decreases with
time\cite{Watson06,Gendre06,Campana06b} as predicted if the absorber
lies at close distance to the GRB and is progressively photoionized by
the burst photons\cite{Lazzati01,Lazzati02b} (see Fig.~\ref{fig:nh}). 

The detection of a variable X-ray column allows us to derive two
important properties of the absorbing medium. First, the fact that the
X-ray column scales almost linearly with metllicity allows us to place
a robust lower limit on the metallicity of the GRB environment. If the
GRB photons would have propagated through a Thompson thick cloud, they
would have been multiply deflected, and the variability pattern would
have been smeared out. Since this is not seen in the observations of
the GRB prompt emission, it is possible to conclude that the
metallicity of the absorber was at least
$Z_{\rm{absorber}}\ge0.03Z_\odot$\cite{Campana06b}.

\begin{figure}
\includegraphics[width=0.8\textwidth]{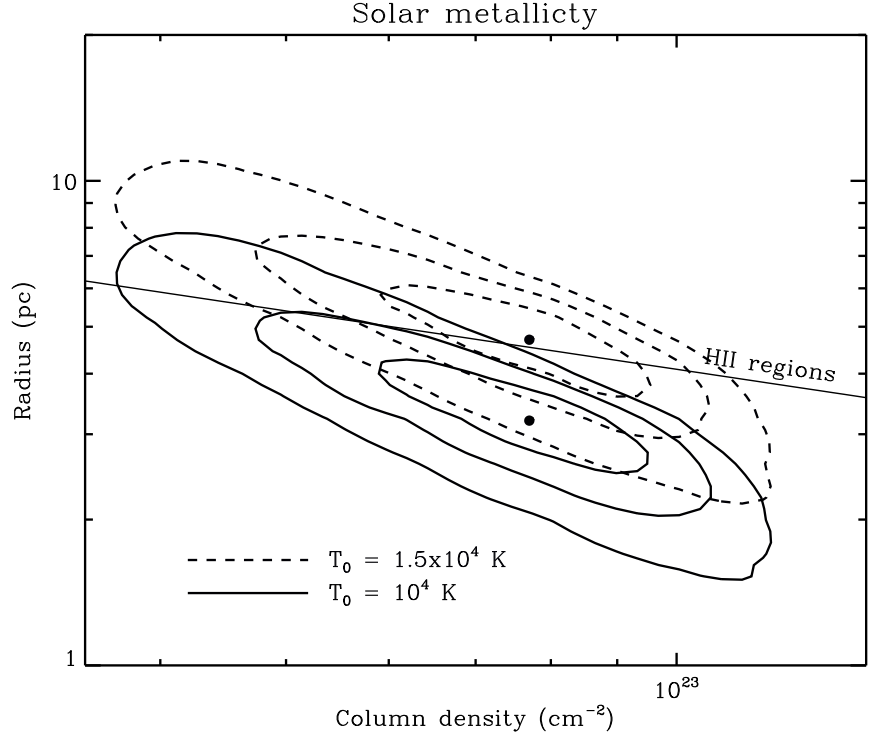}
\caption{Confidence contour (1, 2 and 3-$\sigma$) in the radius --
column density plane for solar metallicity ISM. The computed grid
of radii and column densities over which the fit was performed is much
larger, and spans $10^{16}<R<10^{21}$~cm and
$10^{21}<N_H<10^{25}$~cm$^{-2}$. The straight line shows the locus of
HII regions surrounding massive stars at the end of their life under
the assumption of uniform density of the progenitor molecular
cloud. Despite the simple model, the agreement is
satisfactory. \label{fig:chi}}
\end{figure}

More detailed constraints can be obtained by modelling the evolution
of the observed column density. We used the same time dependent
photoionization code described above\cite{Perna02} to model the
progressive ionization of metals in the ISM and the sublimation of
dust particles. We find that the data can be succesfully reproduced if
we assume that the GRB exploded inside a cavity surrounded by a dense
shell. The shell lies at several parsecs (see Fig.~\ref{fig:chi}) from
the burster and has a total mass of $\sim10^{5}$ solar
masses\cite{Campana06b}.

Interestingly, such geometrical properties are expected from an HII
region surrounding a massive star born inside a dense molecular cloud
(Fig.~\ref{fig:chi}). It is also important to note that due to the
total mass of the absorber, it must be preexisting the GRB
progenitor. In other words, the absorption cannot be produced by the
metals produced by the GRB progenitor. The GRB progenitor was born in
a region that had experienced a previous metal enrichment comparable
to the solar environment. This is an important constraint for GRB
progenitor models, that usually invoke low metallicities in order to
reproduce the amount of angular momentum required by the collapsar
model\cite{Woosley06}.

\section{Summary}

We have analyzed the environment of two well studied gamma-ray bursts,
GRB~021004 and GRB~050904 by means of time dependent
spectroscopy. This technique relays on the fact that an absorber
located at a distance of $\sim1$ to $\sim10$ pc from the burster will
experience progressive ionization on a timescale of seconds to days.
The progressive ionization will be detected observationally as an
evolution of the opacity of the medium. Suitable absorption features
are resonalnt lines (e.g., CIV and SiIV in GRB~021004) and X-ray lines
and continuum (in GRB~050904).

In both the studied events, we find that the environment of the GRB is
consistent with that of a very massive star. What differentiates the
two cases is that the progenitor star evolved in a low-density
environment for GRB~021004, while it evolved in a dense environment in
the case of GRB~050904.

It is a dangerous job to draw conclusions based on only two
cases. This study seem to suggest that the association of GRBs with
massive stars, dramatically proved in low redshift cases, holds up to
very high redshifts. More cases with high quality data have to be
investigated before we can say a final word. We showed that the
analysis of time resolved spectroscopy is a very robust mean of
exploring the GRB environment.


\begin{theacknowledgments}
I am indebted to all my collaborators who made this research possible,
and in particular to R. Perna, S. Campana, F. Fiore and
V. Dwarkadas. This work was partly supported through NASA
Astrophysical Theory Grant NNG06GI06G, NSF grant AST-0307502 and AST
0507571, and {\em Swift} Guest Investigator Program NNX06AB69G and
NNG05GH55G
\end{theacknowledgments}

\end{document}